\documentclass[fleqn,usenatbib]{mnras}

\usepackage{newtxtext,newtxmath}

\usepackage[T1]{fontenc}
\usepackage{ae,aecompl}

\usepackage{graphicx}	
\usepackage{amsmath}	
\usepackage{amssymb}	
\usepackage{bm}         
\usepackage{gensymb}    

\newcommand{\rmin}{r_\mathrm{min}}
\newcommand{\rmax}{r_\mathrm{max}}

\title[Basis for spherical shells]{Proper Fourier decomposition formalism for
cosmological fields in spherical shells}

\author[Lado Samushia]{
Lado Samushia\thanks{E-mail: lado@k-state.edu}
\\
Department of Physics, Kansas State University, 116 Cardwell Hall,
Manhattan, KS, 66506, USA\\
National Abastumani Astrophysical Observatory, Ilia State University,
2A Kazbegi Ave., GE-1060 Tbilisi, Georgia\\
}

\pubyear{2019}

\begin{document}
\label{firstpage}
\pagerange{\pageref{firstpage}--\pageref{lastpage}}
\maketitle

\begin{abstract}
Cosmological random fields are often analysed in spherical Fourier-Bessel
basis. Compared to the Cartesian Fourier basis this has an advantage of
properly taking into account some of the relevant physical processes
(redshift-space distortions, redshift evolution). The observations usually come
in redshift slices and have a partial sky coverage. These masking effects
strongly correlate Fourier-Bessel modes that are meant for a perfect spherical
geometry and result in a lot of redundant measurements. This work proposes a
new Fourier basis that is better suited for measurements in redshift shells and
results in fewer Fourier modes, with the radial modes strictly uncorrelated on
large scales and the angular modes with significantly reduced redundancy. I
argue that the spherical Fourier analysis of cosmological fields should always
use these new modes instead of the historically established Fourier-Bessel
eigenfunctions. The new angular modes on the other hand have number of
practical advantages and disadvantages and whether or not to adopt them for a
particular analysis should be made on a case by case basis.
\end{abstract}




\section{Introduction}

Cosmological fields are often analyzed in Fourier basis, 
\begin{equation}
\label{eq:Finfinity}
f(\bm{r})\propto \displaystyle\int\!\!\mathrm{d}\bm{r}
\tilde{f}(\bm{k})e^{i\bm{k}\bm{r}}.
\end{equation}
\noindent
This is convenient because the distribution of matter in the Universe is
initially very close to Gaussian with different Fourier modes statistically
independent form each other. Although, gravitational instability couples
initially independent Fourier modes on smaller scales, on large scales, where
the gravitational evolution is close to linear, they remain uncorrelated.

In practice, these cosmological fields are observed in finite volumes. The ``
window'' effects due to finite volume correlate the measured Fourier modes
$\tilde{f}(\bm{k})$ and reduce the number of effectively uncorrelated modes
\citep{1991MNRAS.253..307P}.  The exact nature of these uncorrelated modes
depends on the geometry of the observed volume.  For a cube with a side length
$L$ and periodic boundary conditions the proper decomposition into complete
orthogonal basis vectors is be given by,
\begin{equation}
\label{eq:Fcube}
f(x,y,z)\propto
\displaystyle\sum_{\ell,m,n}\tilde{f}_{\ell m n}e^{i\frac{2\pi}{L}\left(\ell x
+ my + nz\right)},
\end{equation}
\noindent
where $l$, $m$, $n$ are integers and $x$, $y$, $z$ are Cartesian coordinates.
The new, smaller set of eigenvectors $\tilde{f}_{\ell m n}$ are now
uncorrelated. The original modes from Eq.~\eqref{eq:Finfinity} are not
independent anymore and can be expressed in terms of the basis of
Eq.~\eqref{eq:Fcube}.

For a sphere of radius $R$, with zero boundary conditions on the outer
boundary and a restriction that the functions are finite everywhere inside, the
complete orthogonal basis is
\begin{equation}
\label{eq:Fsphere}
f(r,\theta,\phi) \propto \displaystyle\sum_{\ell,m,n} \tilde{f}_{\ell m n}j_\ell\left(\frac{k_{\ell n}}
{R}r\right)Y^\ell_m
(\theta,\phi).
\end{equation}
\noindent
Here $j_\ell$ are spherical Bessel functions of the first kind, $Y^\ell_m$ are
the spherical harmonics, and $k_{\ell n}$ is the n-th zero of the $\ell$-th
order spherical Bessel function. $r$, $\theta$, $\phi$ are spherical
coordinates, while $n$, $\ell$ and $m$ are integers satisfying $-\ell \leq m
\leq \ell$ \citep[see e.g.,][]{1995MNRAS.275..483H}.

Real survey volumes are neither cubes nor spheres and have a rather complicated
geometry. A standard practice is to embed these volumes into a larger cube,
zero-pad the areas outside the actual observable volume, and Fourier decompose
the field in the basis of Eq.~(\ref{eq:Fcube})
\citep{1994ApJ...426...23F,1995ApJ...455..429T}.  Alternatively, one can use an
enclosing sphere and the basis of Eq.~(\ref{eq:Fsphere})
\citep{2012A&A...540A..60L}. 

The Cartesian basis is more convenient to compute and interpret and is more
frequently used in the analysis of galaxy clustering data. The spherical basis
has the advantage that some of the relevant physics (e.g. redshift-space
distortions, redshift evolution) is better aligned with the coordinates and
therefore the wide-angle effects can be taken into account more naturally
\citep{2015MNRAS.452.3704S,2015MNRAS.447.1789Y}.

In both cases we are using a larger than observed volume for Fourier
decomposition which gives rise to so called ``Window effects''
\citep[see e.g.,][]{2017MNRAS.464.3121W}. The basis vectors are not orthogonal anymore.
Since the decomposition volume is larger than the observed volume there are too
many basis vectors and not all of them are independent. This affects large
scales (comparable to the size of the survey) more and correlates those modes
even when they are Gaussian. If we decomposed the distribution in true
eigenvectors of the observed volume large scale linear modes would be
uncorrelated. Unfortunately, this is not possible due to a complicated nature
of the visibility masks.

The basis proposed in this paper comes very close to this goal. The method
relies on the fact that surveys are usually analyzed in redshift slices with a
sharp boundary at $\rmin$ and $\rmax$ away from the observer, forming a
fraction of a spherical shell. The \textbf{exact} eigenvectors of a spherical
shell can be easily computed and expressed in terms of a mixture of spherical
Bessel functions of first and second kinds. We can go one step further and
enclose the observed angular mask into a bounding spherical cap. The exact
eigenvectors for a spherical cap can also be computed and expressed in terms of
Legendre functions of the first and second kind. This results in a new basis
for Fourier decomposition that is similar to the spherical basis but has fewer
(less correlated) modes. The basis can be described by $k_n$ - a wavenumber of
a radial mode, and two orbital numbers $\lambda$ and $m$ that describe angular
patterns. Unlike spherical Fourier decomposition $ \lambda$ is not in general 
an integer. This radial and angular eigenvectors have been separately
studied before in the context of geophysics
\citep{1985JGR....90.2583H,2006JGRB..111.1102T,Mushref2010}, but I am not aware
of any reference that studies them simultaneously. I am also not aware of any
works that implement the formalism in the context of cosmological random
fields.

The proposed basis has a number of advantages, especially when describing
clustering on large scales. The radial modes are in fact exact and fully
uncorrelated. Since survey masks do not form a perfect spherical cap, the
angular modes are not exact but are closer to the ``true'' basis than the
conventional spherical harmonics decomposition. As a result the new basis has
fewer eigenvectors for the same range of scales that are much less correlated
because of the reduced redundancy.

The new basis is not computationally more expensive than the standard spherical
decompositions and may be advantageous when analyzing large-scale structure on
large (compared to the size of the sample) scales such as the analysis of
non-Gaussianity in galaxy and weak lensing surveys, and intensity mapping
experiments
\citep[e.g.,][]{2016ApJ...833..242L,2014MNRAS.442.1326K}.

I will derive this basis in section~\ref{sec:basis}, will describe some of
its basic properties in section~\ref{sec:properties}, and possible applications
in section~\ref{sec:conclusions}. Application of the basis to real data is left
for future work \citep[see,][for an example of how the new basis is used in
computing signal to noise of 21cm experiments]{Pullen2019}.

\section{Complete Basis for a Spherical Shell}
\label{sec:basis}

\subsection{Perfect Spherical Shell}
\label{ssec:perfectshell}

I will start by finding a complete basis for a field in a spherical shell
between $\rmin < r < \rmax$, for now assuming a full-sky coverage. The most
general solution of Laplace equation in spherical coordinates is given by

\begin{align} 
\label{eq:generalsol}
f(r,\theta,\phi) = &\left[C_jj_\lambda(kr) + C_yy_\lambda(kr)\right]\times\\
\nonumber
&\left [C_pP^\mu_\lambda (\cos (\theta)) + C_qQ^\mu_\lambda(\cos
(\theta))\right]\times\\
\nonumber
&\left[C_+e^{i\mu\phi} + C_-e^ {-i\mu\phi}\right],
\end{align} 
\noindent
where all $C$s are constants, $Q^\mu_\lambda$ and $P^\mu_\lambda$ are Legendre
functions of the first and second kind, and $j_\ell$ and $y_\ell$ are
spherical Bessel functions of first and second kind.

The quantization of the angular part is standard. First of all, the function
has to be $2\pi$ periodic in azimuthal angle $\phi$, which forces $\mu$ to be
an integer $\mu = m = 0, 1, 2, \ldots$. The functions also have to be finite
for $-1 \leq \cos(\theta) \leq 1$, which in addition requires that $C_q=0$ and
$\lambda$ is a positive integer $\lambda = \ell = 0, 1, 2, \ldots$ and that
$-\ell \leq m \leq \ell$.  For the angular part so far we have regular
spherical harmonics.

We now want to select a subset of radial functions that satisfy necessary
boundary conditions at the edges of the spherical shell. Existing literature
seems to prefer the Newman boundary condition prescribing zero normal derivatives at
these boundaries and I will follow this tradition, although in reality using
Dirichlet or mixed boundary conditions would not make a big practical
difference for large scale modes (except for a Gibb's phenomenon close to the
boundary).\footnote{Periodic boundary conditions do not make much sense for a
radial coordinate.} This means that we have to find a solution to a pair of
equations
\begin{align}
\label{eq:bc1}
j'_\ell(k\rmin) + \frac{C_y}{C_j}y'_\ell(k\rmin) = 0,\\
\label{eq:bc2}
j'_\ell(k\rmax) + \frac{C_y}{C_j}y'_\ell(k\rmax) = 0. 
\end{align} 
\noindent 
These equations will only have a joint solution for specific discrete values of
k and the $C_y/C_j$ ratio. These pairs of values are easy to find numerically
and can be pre-tabulated for arbitrary index $\ell$, and the values of $\rmin$
and $\rmax$. I will denote these solutions by $k_{\ell n}$ and $A_{\ell n}
\equiv C_y/C_j$ where n indexes the solutions of Eqs.~(\ref{eq:bc1}) and
(\ref{eq:bc2}). The new radial functions are then
\begin{equation} 
\label{eq:newbasis}
\mathcal{J}_{\ell n}(r) = j_\ell(k_{\ell n}r) + A_{\ell n}y_\ell(k_{\ell n}r),
\end{equation}
\noindent
and we end up with a new basis
\begin{equation}
\mathcal{F}_{n \ell m}(r,\theta,\phi) \propto \mathcal{J}_{\ell n}(r)Y^m_\ell(\theta,\phi).
\end{equation}
Any function in a spherical shell can be decomposed into this basis by
\begin{equation}
f(r,\theta,\phi) = \displaystyle\sum_{n,\ell,m}\widetilde{f}_{n \ell m}\mathcal{J}_
{\ell n}(r)Y^m_\ell(\theta,\phi),
\end{equation}
\noindent
where $\widetilde{f}_{n \ell m}$ are the new Fourier coefficients. Real data is
never given on a full-sky and therefore the angular large-scale modes will
still be correlated by the mask, but purely radial modes are now strictly
uncorrelated.

\subsection{Reducing Angular Correlations}

When the angular footprint is not a full sky (which is always the case) the
true eigenvectors are not traditional spherical harmonics. For arbitrary
angular masks there is little hope of finding analytically traceable solutions,
for simple geometries however such solutions may exist. One example of such a
geometry is a spherical cap given by $\cos(\theta_\mathrm{max}) < \cos(\theta) <
1$, where we directed azimuthal axis towards the middle of the cap for
simplicity. In this case the eigenvectors are still given by a mixture of
$P^\mu_\lambda$ and $Q^\mu_\lambda$ but the boundary condition changes to the
angular derivative being zero at $\theta = \theta_\mathrm{max}$ and the
functions being finite within $\theta_\mathrm{max} < \theta < 0$. This excludes
$Q^\mu_\lambda$ which are not finite for $\cos(\theta) < 1$ and quantizes
$\lambda$ values in $P^\mu_\lambda$ which are still discrete but not in general
integer. Since we need the solution to be azimuthally symmetric $\mu = m$ still
needs to be integer. 

To find the spectrum in $\lambda$ we have to numerically solve equation
\begin{equation}
\frac{\mathrm{d}P^m_\lambda(\theta)}{\mathrm{d}\theta}\Bigr|_{\substack{\theta=\theta_\mathrm{max}}}
= 0.
\end{equation}
\noindent
$k_{\ell n}$ spectrum for the spherical Bessel functions can then be found as
before by solving Eqs.~\eqref{eq:bc1}--\eqref{eq:bc2} but with $\lambda$s now
being non-integer and determined by the size of the angular cap.

I provide a jupyter-notebook that demonstrates how these spectra can be found
numerically.\footnote{\url{https://github.com/ladosamushia/Spherical-basis-for-a-shell/blob/master/SFBonshell.ipynb.}}

\section{Properties of the New Basis Functions}
\label{sec:properties}

\begin{figure}
\includegraphics[clip,trim={0 1.2cm 0 1.5cm}, width=0.5\textwidth]{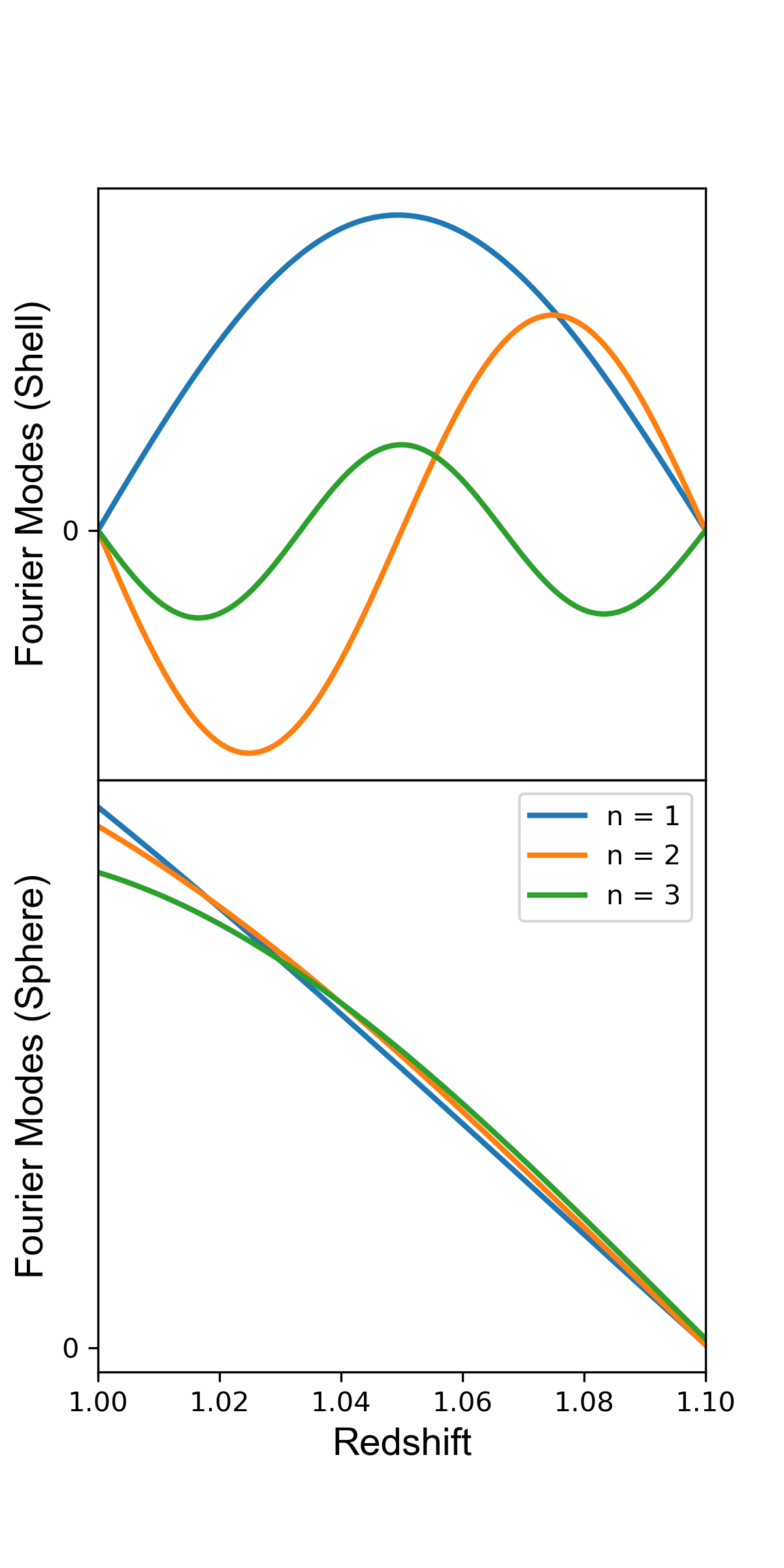}
\caption{First three basis functions (for $\ell = 0$) for the new basis
advocated in this paper (top panel) and the conventional Fourier-Bessel basis
(bottom panel) for a redshift range between $z_\mathrm{min} = 1$ and
$z_\mathrm{max} = 1.1$.}
\label{fig:radmods}
\end{figure}

Radial basis functions described in previous section behave as one would expect
from an orthogonal set. The $n^\mathrm{th}$ basis function has $n - 1$ nodes,
and higher orders pick up information from smaller scales. The top panel of
figure~\ref{fig:radmods} shows first three basis functions for a spherical
shell between $z_\mathrm{min} = 1$ and $z_\mathrm{max} = 1.1$ at $\ell = 0$.
These eigenfunctions are clearly uncorrelated unlike the first three basis
functions of the standard Fourier-Bessel decomposition shown on the bottom
panel of the same figure which are clearly redundant.\footnote{For simplicity,
in this section I compute eigenfunctions to be zero at the boundaries rather
than to have a zero derivative. This does not really affect any of the results
but makes plots easier to interpret.}

\begin{figure*}
\includegraphics[clip,trim={0 1cm 0 1cm},width=\textwidth]{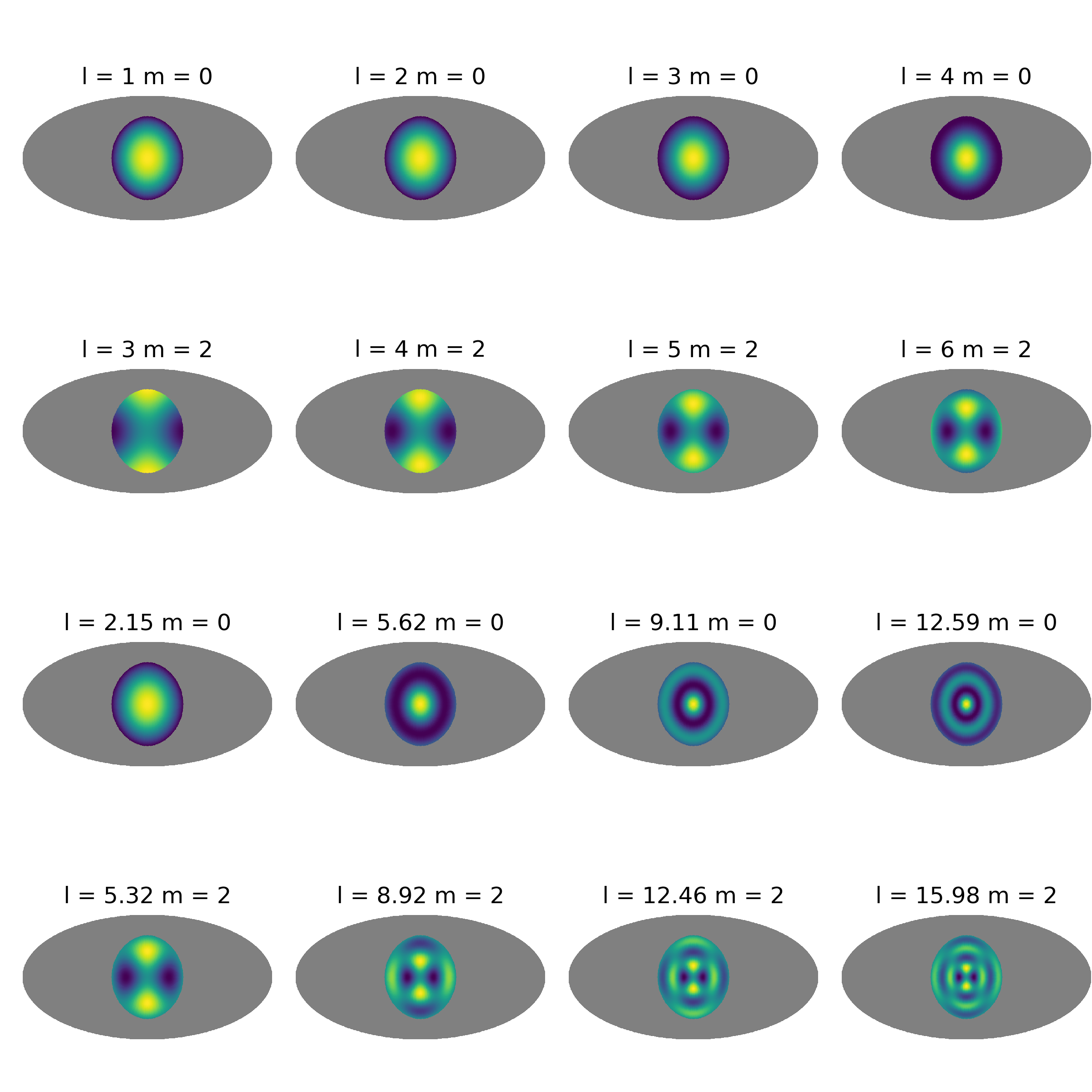}
\caption{First few eigenfunctions for the standard spherical harmonic basis
(top two rows) and the new basis (bottom two rows) for a spherical cap with
$\theta_\mathrm{max} = 52\degree$.}
\label{fig:angmodes}
\end{figure*}

Figure~\ref{fig:angmodes} shows angular patterns produced by the standard
spherical harmonics (top two rows) and by the proper angular basis (bottom two
rows) in a spherical cap with $\theta_\mathrm{max} = 52\degree$ \citep[roughly
the size of the BOSS CMASS North footprint,][]{2013AJ....145...10D}. It is clear
that the spherical harmonics are tightly correlated while the proper basis
functions display the expected pattern of nodes with increasing $\lambda$.

\begin{figure}
\includegraphics[clip,trim={0 0.5cm 0 0.5cm},width=0.5\textwidth]{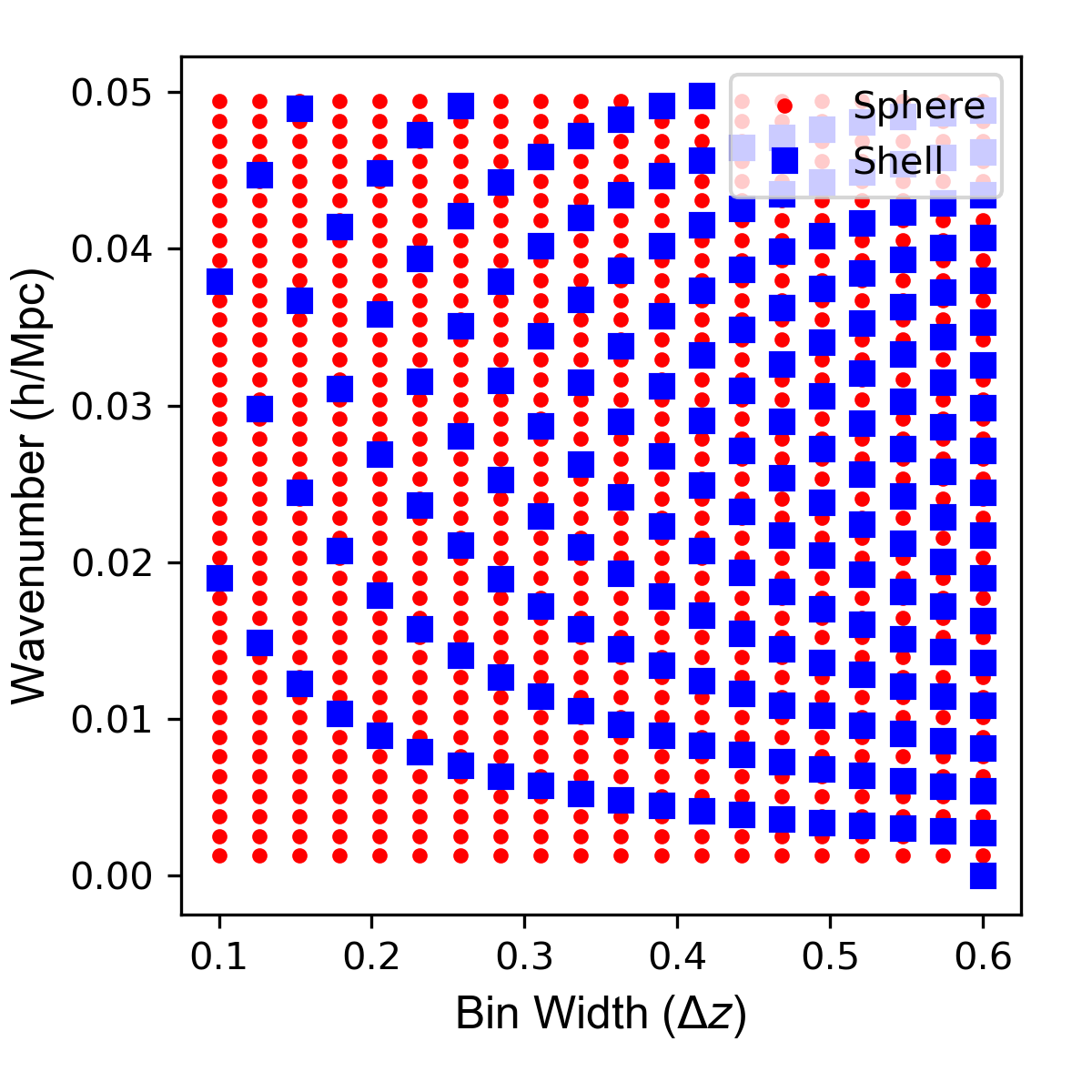}
\caption{Frequencies of the basis functions for the new basis (blue squares)
and the conventional Fourier-Bessel basis (red dots) for a redshift slice with
the upper edge at $z_\mathrm{max} = 1$ and as a function of the bin width.}
\label{fig:radmodesnum}
\end{figure}

Figure~\ref{fig:radmodesnum} shows how the number of radial eigenfunctions
depends on the width of the redshift bin. The upper edge of the redshift slice
is fixed at $z_\mathrm{max} = 1.1$ while the bin width varies. For narrower
bins the standard Fourier-Bessel provide way too many large scale modes that
are bound to be strongly correlated by the radial mask (this spectrum only
depends on the upper edge and not on the bin width), while the new basis has
fewer eigenfunctions for narrow bins properly reflecting this fact. The lowest
frequency modes scale roughly as $\pi/\Delta r$ with bin width as expected.
Also as expected, as the bin width increases (higher frequencies) the
differences between two bases become smaller.

\begin{figure}
\includegraphics[clip, trim={0 0.5cm 0 0.5cm},width=0.5\textwidth]{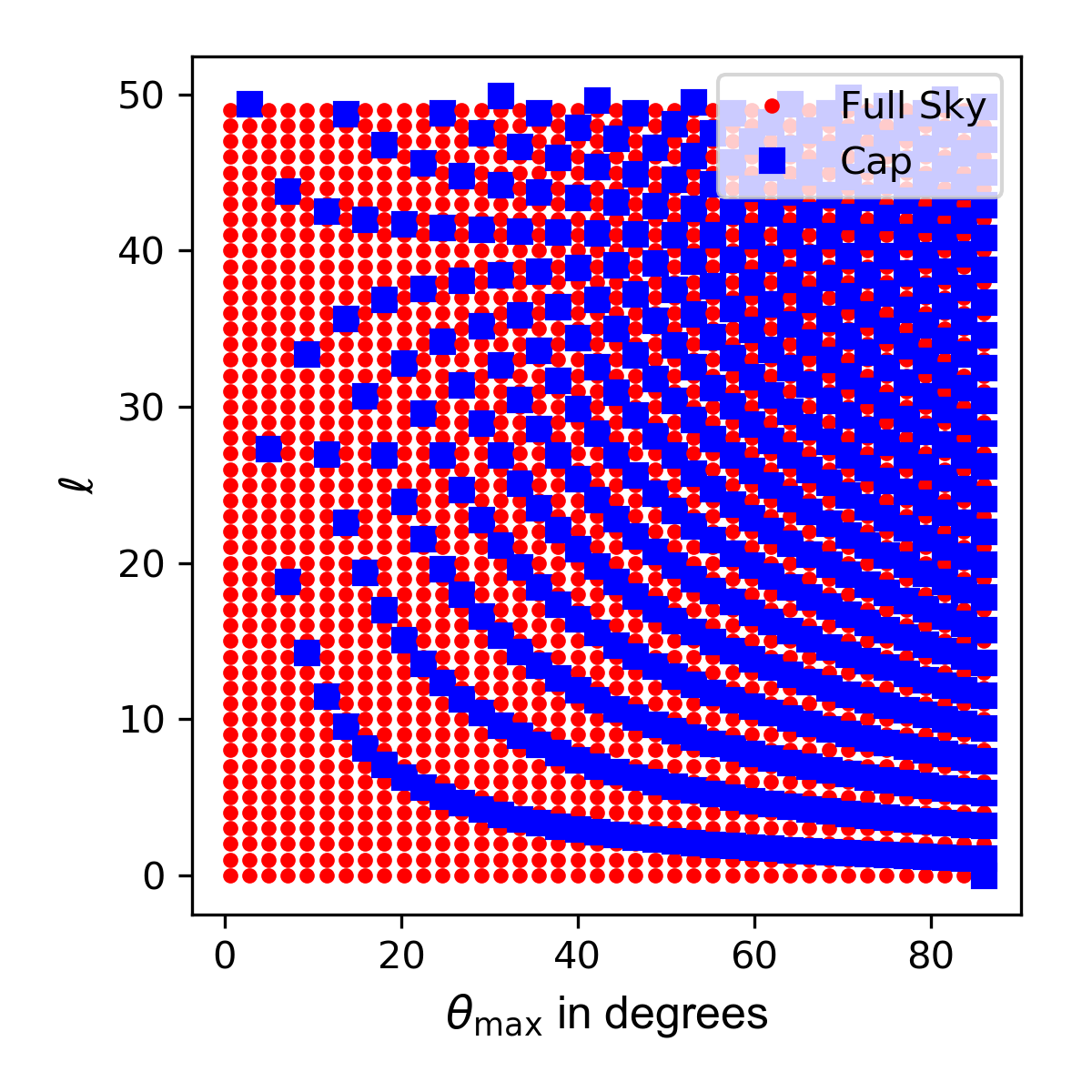}
\caption{$\lambda$ spectrum of proper angular basis functions for a spherical
cap (blue squares) versus the standard spherical harmonic $\ell$ spectrum (red
dots) as a function of $\theta_\mathrm{max}$.}
\label{fig:angmodesnum}
\end{figure}

Figure~\ref{fig:angmodesnum} shows a similar mode count for the angular
eigenfunctions as a function of the footprint size. For the standard spherical
harmonics basis the $\ell$ spectrum consists of positive integers irrespective
of the size of the footprint which results in strongly correlated modes at low
$\ell$. The proper basis has fewer eigenfunctions for smaller footprints. The
lowest $\lambda$ roughly scales as $\pi/\theta_\mathrm{max}$ as expected. For
larger footprints the difference between two bases gets smaller. The $\lambda$
spectrum has all the intuitively expected properties e.g. for half-sky coverage
it consists of only odd integers (or even integers depending on the boundary
conditions) or half the original full sky modes.

\section{Conclusions}
\label{sec:conclusions}

This paper introduces a new Fourier basis that is exact for fields observed in
spherical shells and caps. This basis has a number of potential advantages.  It
strongly reduces the window induced correlations between measured Fourier
modes, in fact if the survey geometry were a perfect spherical shell in an
angular cap there would be no correlations. Realistic survey footprints are
unlikely to be exact spherical caps but one can always find the smallest
spherical cap that encloses the survey footprint (e.g. BOSS CMASS North can be
enclosed in a spherical cap with $\theta_\mathrm{max} \simeq 52\degree$). This
will not completely remove the correlations but will reduce them significantly.
The window functions are not always easy to treat, especially on very large
scales, and the studies that make use of very large scale power spectrum may
want to opt for the Fourier coefficients that are significantly decorrelated.
The radial part of the decomposition of course does not suffer from this
problem since observed fields are almost always analysed in redshift shells
with sharp radial boundaries. Since significantly less Fourier modes need
to be computed the analysis codes are likely to be much faster, and the
covariance matrices are smaller which is always a welcome improvement.

For the standard BAO/RSD type of analysis that takes place on scales smaller
than the survey window the Cartesian Fourier basis is probably more convenient,
especially given recently developed methods that reduce wide-angle effects
\citep{2015MNRAS.453L..11B,2015PhRvD..92h3532S}. For the analysis of very large
scale physics (e.g. primordial non-Gaussianity, relativistic effects, etc.) the
spherical basis may be preferable. For those types of studies one should use
the new radial shell eigenvectors of Eq.~\eqref{eq:newbasis} instead of
historically used spherical Fourier-Bessel basis.  They provide uncorrelated
modes and carry no additional numerical or conceptual penalty. Whether or not
to also use the spherical cap basis is a more complicated question. Even for non
full-sky footprints one may opt to go with the standard spherical harmonics
decomposition for a number of reasons (theoretical familiarity, the
availability of well validated codebase for computing $C_\ell$s, the fact that window
effects are not going to completely go away anyway). A good practical
compromise in this case may be to find the $\lambda$ spectrum for the enclosing
spherical cap and than only use the standard spherical harmonics eigenvectors
with the nearest integer $\ell$ values.

\section*{Acknowledgements}
I would like to thank Yu Hai and Anthony Pullen for useful discussions. This
work was funded by NASA grant 12-EUCLID11-0004 and the DOE grant DE-SC001184.








\appendix


\bsp	
\label{lastpage}
\end{document}